# Spatial and Temporal Resolution in Entangled Ghost Imaging

J. Reintjes and Mark Bashkansky
KeyW Corp, Hanover, MD

and

Mark Bashkansky
Code 5613, Naval Research Laboratory, Washington, DC

## ABSTRACT

We show that, when the integration time of the single photon detectors is longer than the correlation time of the biphoton, the attainable spatial resolution in ghost imaging with entangled signal idler pairs generated in type II spontaneous parametric down conversion is limited by the angular spread of single-frequency-signal idler pairs. If, however, the detector integration time is shorter than the biphoton correlation time, the transverse k-vectors of different spectral components combine coherently in the image, improving the spatial resolution.

## 1. INTRODUCTION

We discuss factors affecting spatial resolution in ghost imaging using entangled signal/idler photons generated in spontaneous parametric down conversion (SPDC). We show that, for conditions commonly encountered in experiments, the spatial resolution is much poorer in the e-direction than in the o-direction. This behavior is due to the angular dependence of the extraordinary refractive index in the SPDC crystal, which leads to a greater restriction on the angular width of the phase matching peak. We further show that, if the time resolution of the single photon detectors is fast enough to resolve the coherence time of the biphoton, the contribution of different wavelength components can combine coherently in the image, improving the spatial resolution.

## 2. ENTANGLED GHOST IMAGING

We consider the entangled ghost imaging geometry shown in Fig. 1. The entangled signal idler pairs are generated through SPDC in a 3 mm long BBO crystal using type II phasematching in the forward direction. The idler photon is separated with a polarizing beamsplitter and propagates through an imaging lens L1 and a physical slit. Those idler photons that pass through the slit are collected with lens L2 and detected with photo detector D1. Signal photons propagate forward from the SPDC crystal to the ghost image plane where the spatial distribution is detected in temporal correlation with the detected idler photons using a scanning or spatial resolving single photon detector D2. The formation of the virtual image in this geometry has been discussed by several authors [1-5]. General discussions of parameters affecting the image formation and spatial resolution have been discussed in [2]. Here we discuss in more detail those factors that affect spatial resolution.

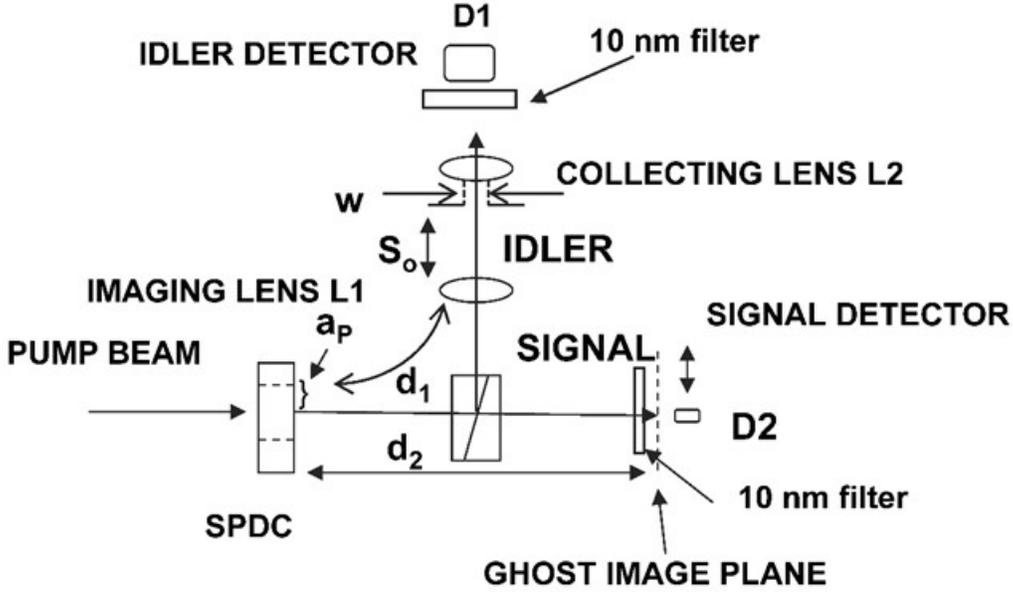

Fig. 1. Schematic diagram of entangled ghost imaging

It has been demonstrated by several authors [1,2] that the ghost image obtained in correlation is described by combinations of the same diffraction integrals that describe first order classical imaging. In particular, the resolution of the image is determined by the range of effective spatial frequencies contained in the signal and idler photons and by the effective numerical aperture of the imaging system. For the diagram in Fig. 1, the effective numerical aperture is given as the ratio of the pump beam radius to the distance from the crystal to the virtual image plane:

$$NA = \frac{a_p}{d_2}. \qquad (1)$$

Unlike in conventional classical imaging, the range of spatial frequencies available for forming the image is not determined by the dimensions or features of the object (here the physical slit in the idler path) but rather by the angular phase matching width of the SPDC interaction. An additional difference between conventional imaging and the correlated image configuration of Fig. 1 is that, in conventional imaging when the illuminating signal is polychromatic, each spectral component independently acquires the full range of spatial frequencies necessary for forming an image. In contrast, for correlated ghost imaging with SPDC generated signal and idler photons, different spectral components can contain a different range of spatial frequencies due to off axis phase matching and the finite angular acceptance of the phase matching conditions.

## 2.1 PHASE MATCHING

Phase matching diagrams are given in Fig. 2 for various conditions.

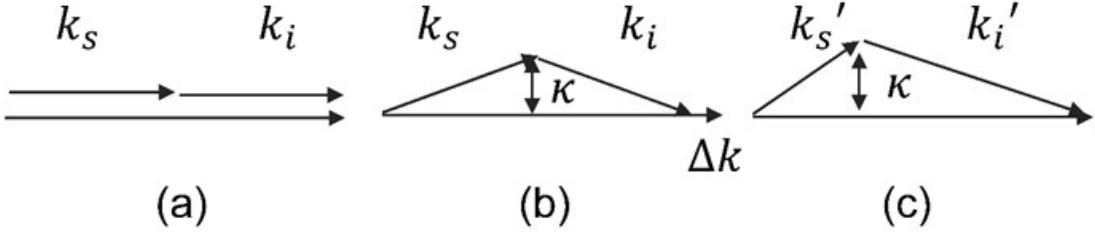

Fig. 2. Phase matching diagrams for SPDC. (a) Degenerate forward phase matching. (b) Forward phase mismatch with off axis beams. (c). Non-degenerate off-axis phase matching

The phase mismatch in the z direction in the SPDC interaction is given by

$$\Delta k_z = k_{pz} - k_{sz} - k_{iz} \tag{2}$$

We choose a crystal orientation for which the SPDC interaction is phase matched in the forward direction for degenerate signal idler pairs. For the degenerate pairs the allowed angular spread is determined by the relation

$$\Delta k_z = k_p - k_{sz}(\theta_s) - k_{iz}(\theta_s) = \frac{2\pi}{L} \tag{3}$$

where L is the length of the crystal.

The phase matching requirement in the transverse direction gives rise to the relation

$$\kappa_P = \kappa_s + \kappa_i, \tag{4}$$

where $\kappa_{P,S,i}$ is the transverse k-vector of the pump, signal and idler, respectively. For the interactions considered here, for which the pump beam does not diffract significantly in the SPDC crystal, we take $\kappa_P = 0$, giving the relation

$$\kappa_s + \kappa_i = 0. \tag{5}$$

For type I phase matching the signal and idler rays are both polarized in the o direction. It is reasonable to assume that the ordinary index of the signal/idler is constant over the narrow range of angles expected in the phase matching spread. In that case the phase mismatch is determined entirely by the reduction of the projection of the signal/idler k-vectors onto the z axis as the angle increases, giving the result

$$\Delta k_z(\theta) = \frac{\kappa_i^2}{2}\left(\frac{1}{k_s} + \frac{1}{k_i}\right) = \frac{\kappa_i^2}{2}\left(\frac{k_s}{k_s^2} + \frac{k_i}{k_i^2}\right) = k_s\theta^2 \tag{6}$$

with a maximum angle

$$\theta_{Imax} = \sqrt{\frac{\lambda}{n^o L}} \qquad (7)$$

For non-degenerate type I phase matching the phase mismatch is generalized to

$$\Delta k_z(\theta) = \frac{\kappa_s^2}{2k_s} + \frac{\kappa_i^2}{2k_i} = \frac{k_s \theta_s^2}{2} + \frac{k_i \theta_i^2}{2} \qquad (8)$$

For the type II interaction used in our experiments one of the generated waves, taken here to be the signal, is polarized in the e direction. The phase mismatch is given by

$$\Delta k_z = -\frac{1}{n_s^e(\Delta\omega,\theta)} \frac{\partial n_s^e(\Delta\omega,\theta)}{\partial \theta} \kappa_s + \frac{\kappa_i^2}{2k_i} + \frac{\kappa_s^2}{2k_s}. \qquad (9)$$

giving rise to the two cones of type II phase matching. For the dispersion constants of BBO, the change in length of the extraordinary polarized idler k vector as the angle increases dominates the k vector mismatch in the near forward direction giving the approximate result

$$\Delta k_z \approx -\frac{1}{n_s^e(\Delta\omega,\theta)} \frac{\partial n_s^e(\Delta\omega,\theta)}{\partial \theta} \kappa_s \qquad (10)$$

with a maximum angle

$$\theta_{IImax} = \frac{1}{\partial n_s^e(\Delta\omega,\theta)/\partial\theta} \frac{\lambda}{L}. \qquad (11)$$

One of the consequences of this result is that the angular spread of the degenerate signal idler pairs in the near forward direction is much narrower for the type II interaction than for the type I interaction.

## 2.2 GHOST IMAGING.

The transverse distribution of the signal at the ghost image plane, $d_2$, detected in correlation with all idler photons that pass through the slit at position $z_{1i}$ for a single frequency pair is

$$CCR(x_{1i}, z_{1i}, x_{2s}, d_2) =$$

$$\int_{-w/2}^{w/2} dx_{1i} \left| \int dx_{os} e^{-\frac{x_{os}^2}{a_p^2}} e^{\frac{i\pi x_{os}^2}{\lambda_s d_2}} e^{-\frac{i2\pi x_{os} x_{2s}}{\lambda_s d_2}} \int d\kappa_i \; sinc\left(\frac{\Delta kL}{2}\right) e^{\frac{i\Delta kL}{2}} e^{-i\kappa_i(x_{os}+x_{1i})} e^{i\frac{\lambda_i d_2}{4\pi}\kappa_i^2} \right|^2 \qquad (12)$$

where w is the width of the slit in the idler and $\Delta k$ is the phase mismatch in the z direction. The point spread function (PSF) for the signal, which can be taken as a measure of the spatial resolution of the system, with a point idler detector at the center of the slit ($x_{1i} = 0$) for a single frequency pair is

$$CCR(0, z_{1i}, x_{2s}, d_2) =$$

$$\left| \int dx_{os} \, e^{-\frac{x_{os}^2}{a_p^2}} e^{\frac{i\pi x_{os}^2}{\lambda_s d_2}} e^{-\frac{i2\pi x_{os} x_{2s}}{\lambda_s d_2}} \int d\kappa_i \, sinc\left(\frac{\Delta kL}{2}\right) e^{\frac{i\Delta kL}{2}} e^{-i\,\kappa_i x_{os}} \, e^{i\frac{\lambda_i d_2}{4\pi}\kappa_i^2} \right|^2 \quad (13)$$

We consider first the ghost image formed for the interaction of Fig. 1 with a slit width of 160 μm and $d_2 = 750 \, mm$ corresponding to a numerical aperture of $NA = .002$. The image formed with single photon correlated counting is shown in Fig. 3a, and the corresponding calculated distribution is shown in Fig. 3b.

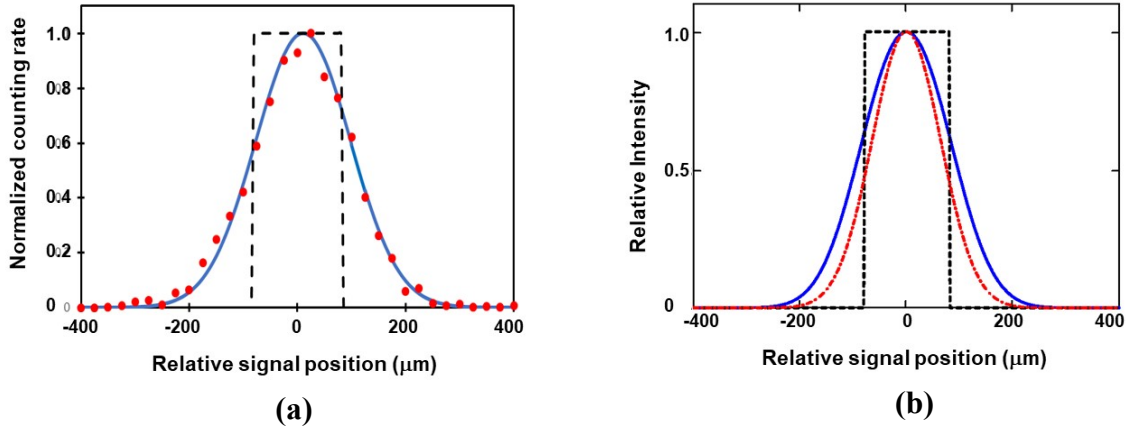

(a)          (b)

Fig. 3. Signal distributions in the ghost image plane for a 160 μm wide idler slit, $d_2$ = 750 mm and NA = .002. (a). Measured single-photon signal distribution (red dots), Gaussian fit to points (solid blue line), equivalent idler slit (dashed black line) (b). Calculated image distribution (solid blue line), Calculated point spread

The results indicate that, while the half width of the measured distribution is comparable to the dimension of the slit, the slit is not resolved. This can be understood from the geometry in Fig. 4 showing that only a limited angular spread in the signal/idler biphoton is effective in contributing to the image under these conditions. For this geometry, the pump beam forms an aperture stop at the crystal, limiting the effective numerical aperture and hence the spatial resolution.

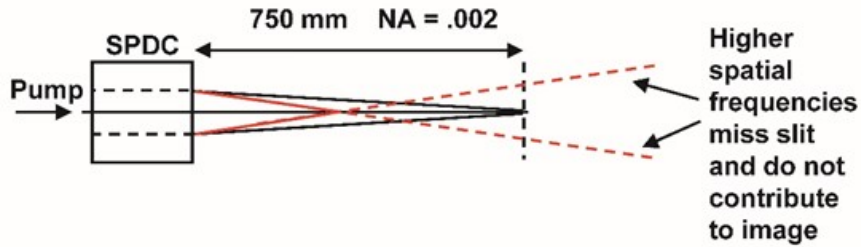

Fig. 4. Geometry for results in Fig 3 showing limited range of effective spatial frequencies because of pump beam aperture. (color online)

This is confirmed in the calculated distribution shown in Fig. 3b under the assumption that all effective spatial frequencies are phase matched. In that case, the expression in Eq. 12 becomes

$$CCR(x_{1i}, z_{1i}, x_{2s}, d_2) = \int_{-\frac{w}{2}}^{\frac{w}{2}} d x_{1i} \left| \int dx_{os} e^{-\frac{x_{os}^2}{a_p^2}} e^{-\frac{i2\pi x_{os} x_{2s}}{\lambda_s d_2}} \right|^2 \quad (14)$$

for the approximation of a single frequency degenerate signal-idler biphoton.

We would then expect that improved spatial resolution would be obtained by moving the crystal closer to the virtual image plane as in Fig. 5.

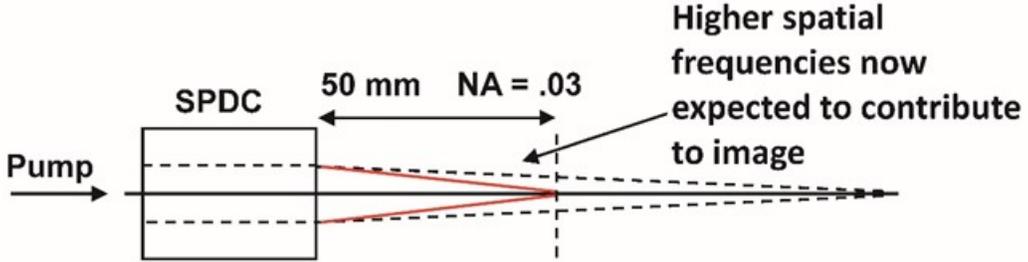

Fig. 5. Imaging geometry with higher numerical aperture and expected wider range of spatial frequencies. (color online)

The expected image as calculated from Eq. 14, assuming all effective rays are completely phase matched, is shown in Fig. 6a, along with the measured correlated image distribution in Fig. 6b for a value of $d_2 = 50\ mm$, corresponding to a numerical aperture of $NA = .03$.

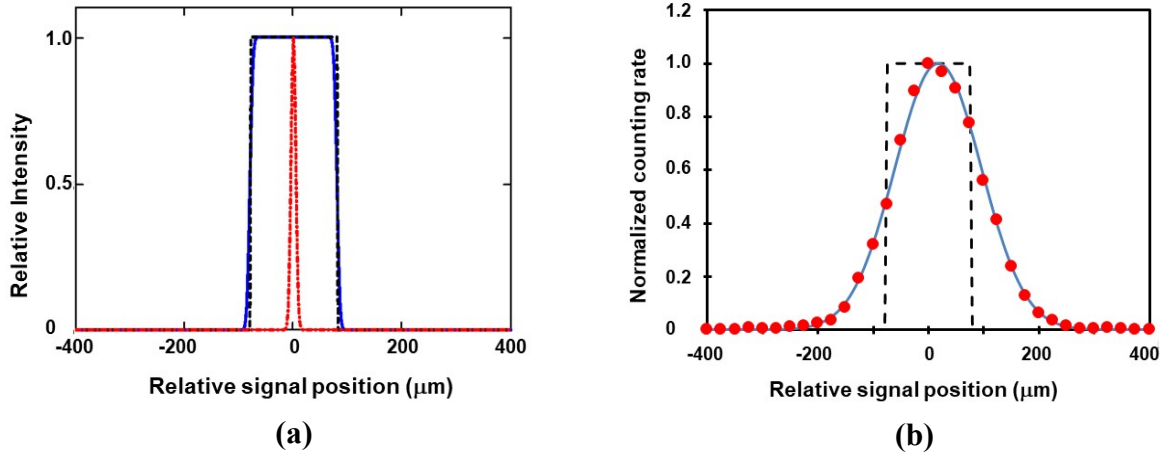

Fig. 6. Signal distributions in the ghost image plane for a 160 µm wide idler slit, $d_2 = 50$ mm and NA = .03. (a) Calculated distribution for geometry assuming all effective rays are phase matched (solid blue curve), calculated point spread function for a point detector at the center of the idler slit (red dash-dot curve), equivalent 160 µm idler slit (black dash curve); (b) Measured single-photon signal distribution (red dots), Gaussian fit to points (solid blue line), equivalent idler slit (dashed black line). (color on line)

For this geometry, the calculation predicts a high-quality reproduction of the 160 μm slit (Fig. 6a), while the measured distribution (Fig. 6b) continues to be a diffraction blur that is little changed from the distribution in Fig. 3a obtained with a numerical aperture of .002.

It has been pointed out [5] that the formulas describing the formation of the entangled ghost image are the same as those that would describe the formation of a classical image at the ghost image plane in a system in which a classical incoherent source illuminates the idler slit from behind and propagates backward along the idler path to the crystal exit face, where it reflects and propagates forward along the signal path to the virtual image plane. We show the image for such a classical measurement using the same optics as in the quantum correlated images and a classical lamp with a 10 nm filter centered at 810 nm in Fig. 7.

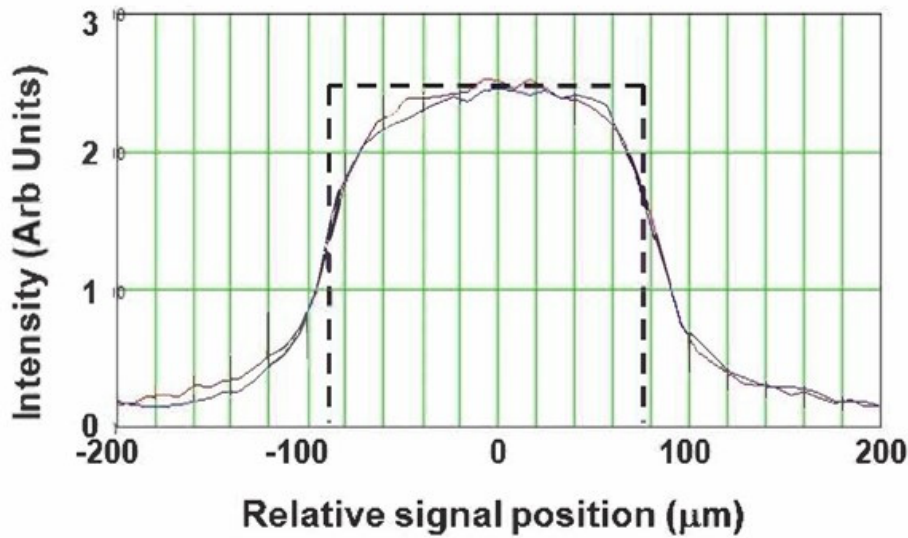

Fig. 7 Image of a classical lamp with a 10 nm filter at 810 nm located behind the 160 μm idler slit and reflecting off the SPDC crystal face. Red curve – direct image; blue curve – image with cylindrical lens to compress slit in vertical direction; black dashed curve – equivalent 160 μm slit. (color on line)

The distribution shows an image of the slit that, while not perfect, has significantly higher resolution than the single photon image of Fig. 6b. The image in Fig 7 also shows the system imaging capability and indicates that the source of limited resolution is intrinsic to the SPDC system and not the experimental limitations. Comparison of Figs 7 and 6b also shows the first report of a discrepancy between the advanced wave model of Ref. 5 and the experimentally observed entangled ghost image.

In order to investigate this behavior further we examined the angular spread of the signal and idler photons produced in the SPDC interaction. The far field distributions for the e and o phase matching cones in a 3 mm long BBO crystal cut for forward phasematching at 405 nm with a 10 nm filter in the signal and idler are shown in Fig. 8. The scale indicates a 10 mrad angular width outside of the crystal.

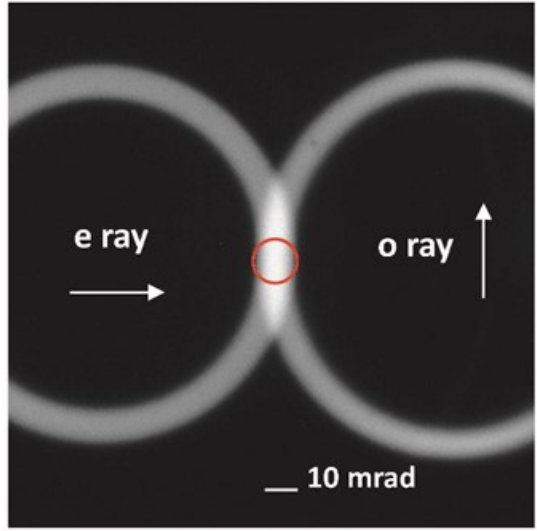

Fig. 8. Far field distribution of radiation from the SPDC crystal showing the e and o cones of type II phase matching. The forward phase matching area of our experiments is indicated by the red circle (color online).

The two cones overlap in the center at the forward phase matching direction indicated by the circle (red online). The angular spread in the overlap region is measured to be 11 mrad, which, while smaller than the 60 mrad spread needed to produce the calculated image of Fig. 6a, could be expected to support a higher resolution than the one measured in Fig. 6b.

### 3. SPECTRAL BANDWIDTH EFFECTS.

We now examine the role of spectral bandwidth, detector temporal resolution and the details of the phase matching angular width in type II phase matching in determining the measured spatial resolution of the ghost image. As discussed in Ref 2, for a polychromatic source with detectors that are too slow to resolve the biphoton coherence time ($t_c \approx 1/spectral\ width$) the contributions of each signal-idler frequency pair combine incoherently in the image as the integral of square magnitudes. For a polychromatic biphoton with slow single photon detectors as used here, the point spread function of the ghost image is

$$CCR(0, z_{1i}, x_{2s}, d_2) = \int d\Delta\omega \left| \int dx_{os}\ e^{-\frac{x_{os}^2}{a_p^2}} e^{\frac{i\pi x_{os}^2}{\lambda_s d_2}} e^{-\frac{i2\pi x_{os} x_{2s}}{\lambda_s d_2}} \int d\kappa_i\ sinc\left(\frac{\Delta k L}{2}\right) e^{\frac{i\Delta k L}{2}} e^{-i\kappa_i x_{os}} e^{i\frac{\lambda_i d_2}{4\pi}\kappa_i^2} \right|^2 \quad (15)$$

where the $\Delta\omega$ integral is taken over the bandwidth of the biphoton.

Fig. 9 shows the SPDC far field again with the angular spread of the single frequency degenerate signal-idler pair indicated as the spacing between the two circles (red online).

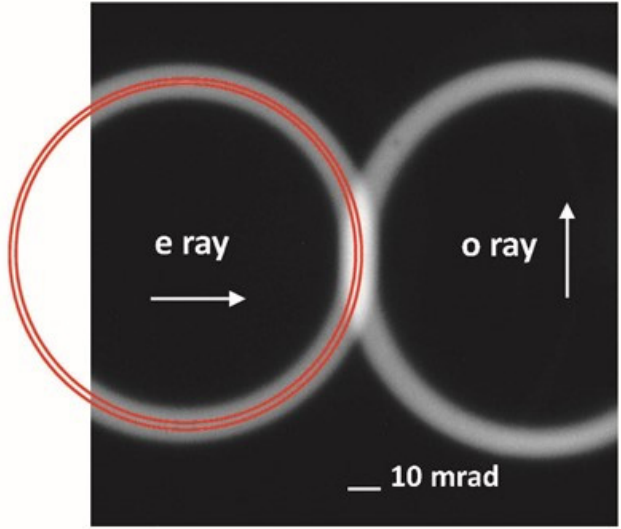

Fig. 9. Far field distribution of radiation from the SPDC crystal. The far field region appropriate for single frequency degenerate operation is between the two solid circles (color online).

The angular width of the single frequency degenerate signal idler pair is measured to be 3.7 mrad, which is comparable to the 4 mrad needed for the small NA image of Fig. 3a. However, this serves now to explain the lack of improvement when the geometry was changed to increase the NA, since the single-frequency SPDC process did not generate a large enough range of transverse spatial frequencies to support the larger NA.

The calculated correlated image for a degenerate signal idler pair is shown in Fig. 10, along with the calculated PSF for diffraction in the e-direction for $d_2 = 50\ mm$. It can be seen that the PSF is of comparable width to the 160 μm idler slit and the correlated image does not resolve the slit, in contrast to the calculated distributions in Fig. 6a.

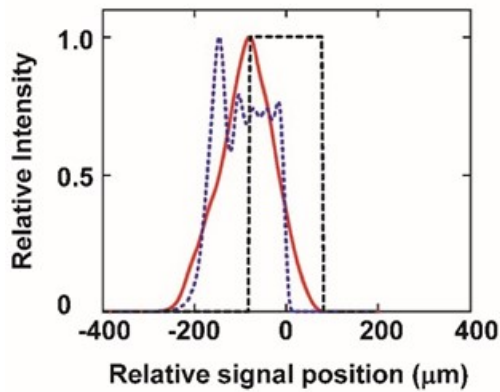

Fig. 10. Calculated image (solid red curve) and PSF (dotted blue curve) for a degenerate signal idler photon pair, a 160 μm slit and $d_2 = 50\ mm$ using the expression in Eq. 9 for $\Delta k$ for diffraction in the e direction. Equivalent idler slit – dashed black curve. (color on line)

We note that the origin of the narrow angular width for the degenerate type II pair in Fig. 9 is the linear term in the phase mismatch of Eq. 9 depending on $\kappa_s$. For the parameters appropriate for type II phase matching in BBO pumped at 405 nm, this term dominates the k-vector mismatch as the angle increases. Essentially it describes a change in length of the signal k-vector due to the variation of the extraordinary index with angle, rather than simply the shortening of the projection of the signal k-vector on the z axis as is the case with type I phase matching. The wider angular spread near the forward direction outside of the single frequency degenerate pair as indicated in Fig. 9 arises from off axis phase matching of non-degenerate frequency pairs within the bandwidth of the 10 nm filter. The implication for imaging resolution is indicated in Fig. 11, which shows the approximate widths of the cones in the e direction for degenerate operation at 810 nm, for an e ray at 805 nm at one edge of the 10 nm filter transmission and for an e ray at 815 nm at the other edge of the filter.

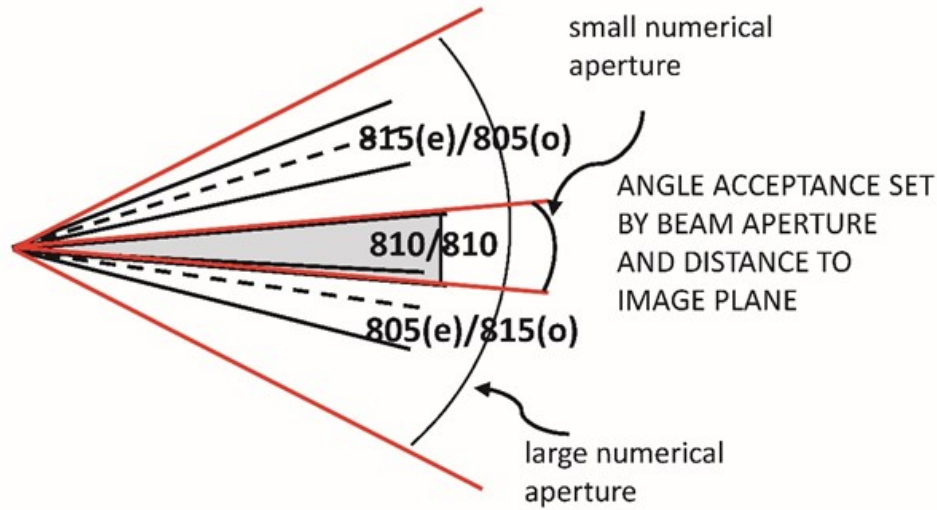

Fig. 11. Schematic illustration of angular extent of type II phase matched generation for different signal-idler frequency combinations within the 10 nm filter bandwidth

While the overall angular spread of photon pairs generated within the 10 nm bandwidth of the filter is large enough to accommodate the imaging of the larger NA, the angular spread of each individual frequency pair is much smaller. In order to take advantage of the wider angular spread of the full bandwidth it would be necessary for the different frequency components to combine coherently in the image. However, as has been pointed out earlier, the different frequency components combine incoherently when the single photon detectors don't resolve the biphoton coherence time. As a result, the image resolution is characteristic of the angular spread of a single frequency pair.

This hypothesis can be tested by noting that the angular spread of a single frequency pair is much greater in the o direction than in the e direction. This results from the different variation of the k vector mismatch in the o-direction than in the e-direction. It can then be expected that the image resolution in the o direction should be better than in the e direction as indicated in Fig. 12 even with the slow detectors used in our experiments.

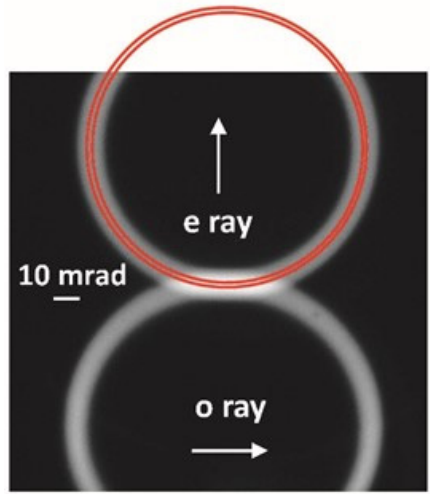

Fig. 12. Far field distribution of radiation from SPDC crystal rotated 90 degrees from the distribution of Fig. 9 to allow measurements to be done with the angular spread in the o direction.

Single photon correlated images obtained with the slit diffracting in the o direction are shown for $d_2 = 50\ mm$ and $w = 160\ \mu m$ in Fig. 13a and $w = 320\ \mu m$ in Fig. 13b.

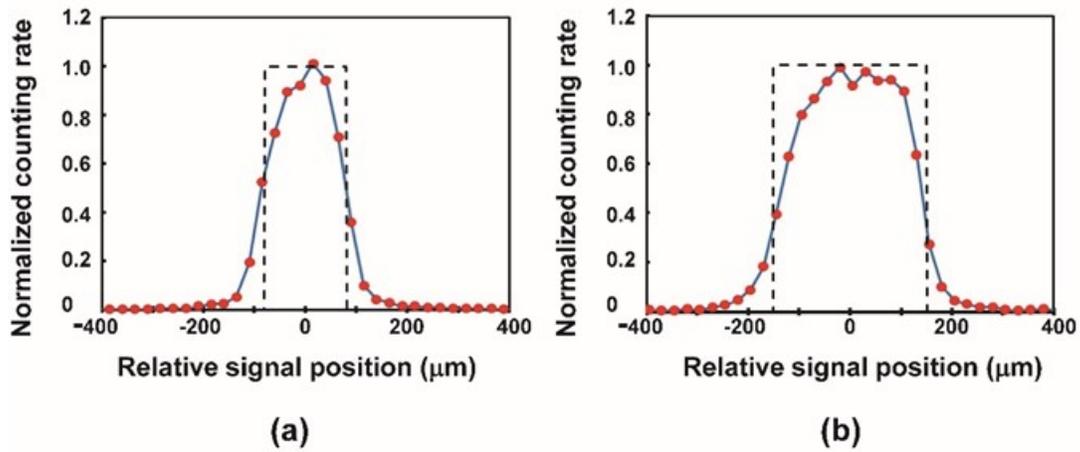

Fig. 13. Measured single-photon signal distributions (red dots) in the ghost image plane for $d_2$=50 mm (NA=.03) using the angular spread in the o-direction. Solid blue curve is drawn through the points. Black dash curve is the equivalent idler slit. (a). 160 μm slit. (b). 320 μm slit. (color online)

The image in Fig. 13a is directly comparable to the one shown in Fig. 6b and shows improved resolution in the form of steeper edges of the slit and narrower wings. The image improvement is illustrated better in Fig. 13b with a wider 320 µm slit. The images in Fig 13, when compared to the one in Fig. 6b, confirm that the spatial resolution in our experiments is dominated by the angular spread of single frequency signal-idler pairs, and is much worse in the e direction than the o direction for type II phase matching.

The explanation of the results in Fig. 6 and 13 is predicated on the incoherent combination of different frequency components in the image resulting from the use of single photon detectors that do not resolve the temporal coherence time of the biphoton. We now consider the use of single photon detectors that are fast enough to resolve the biphoton coherence time. In that case, the amplitudes of the different frequency components combine coherently to form the image. The point spread function is now given by

$$CCR(0, z_{1i}, x_{2s}, d_2) =$$
$$\left| \int d\Delta\omega \int dx_{os} \, e^{-\frac{x_{os}^2}{a_p^2}} e^{\frac{i\pi x_{os}^2}{\lambda_s d_2}} e^{-\frac{i2\pi x_{os} x_{2s}}{\lambda_s d_2}} \int d\kappa_i \, sinc\left(\frac{\Delta kL}{2}\right) e^{\frac{i\Delta kL}{2}} e^{i\frac{\lambda_i d_2}{4\pi}\kappa_i^2} e^{-i\kappa_i x_{os}} \right|^2 \quad (16)$$

The predicted effect of the coherent combination is shown for three crystal lengths in Fig. 14 for $d_2$=50 mm corresponding to NA = .03.

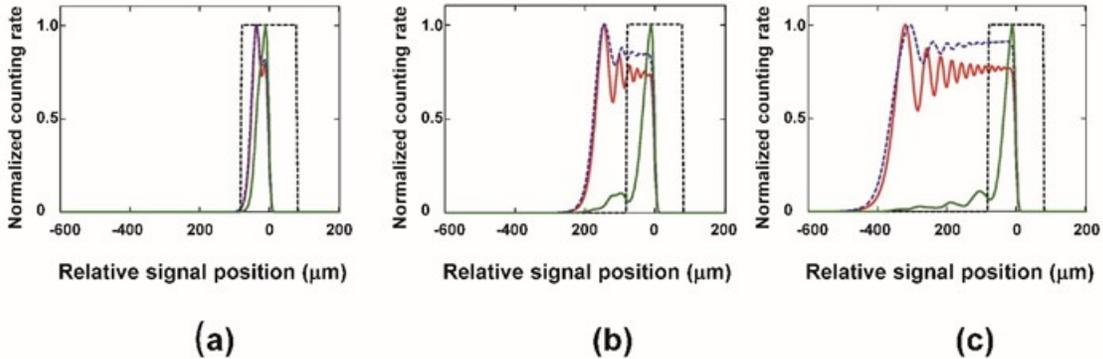

Fig. 14. Calculated point spread functions (PSF) for $d_2 = 50$ mm (NA=.03) and a 160 µm idler slit for incoherent combination of wavelength components in the image (dashed blue curve) and coherent combination (solid green curve). Solid red curve – PSF for single frequency degenerate condition. Equivalent idler slit – black dashed square curve. (a) 1 mm long BBO crystal. (b) 3 mm long BBO crystal (c) 6 mm long BBO crystal. (color online)

For the 3 mm crystal used in our experiments the point spread function for a single frequency degenerate pair, shown as the dashed blue curve in Fig. 14b, is comparable to the width of the 160 µm slit as discussed earlier. The point spread function for the incoherent combination of the slow detector, shown as the dashed blue curve, is comparable to the single frequency PSF, again as illustrated in our experiments. However, the point spread function for the coherent combination of a fast detector, shown as the solid green curve, is considerably narrower and would produce an image with higher resolution. The effect becomes even more pronounced for a 6 mm long crystal (Fig. 14c), while the differential spatial resolution practically disappears for a 1 mm long crystal (Fig 14a). This variation can be understood in terms of the angular spread of the phase matching peaks for the various frequency combinations as shown in Fig. 15.

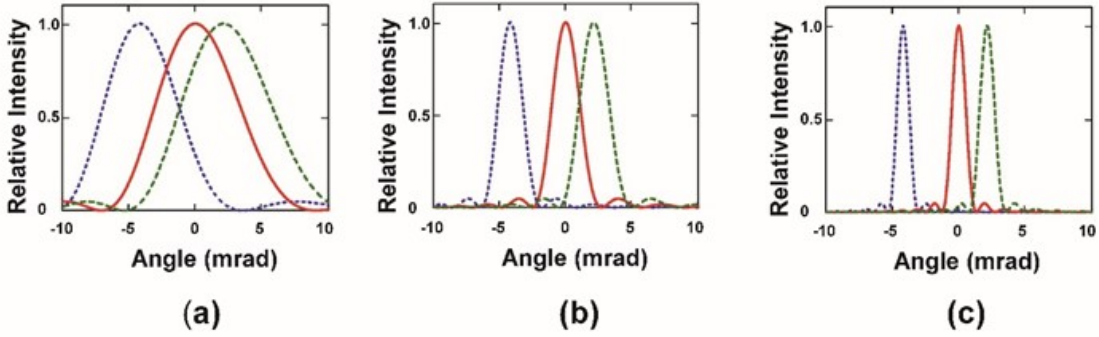

Fig. 15. Phase matching curves vs exterior angle for degenerate operation (solid red curve), signal ray at 815 nm (dotted blue curve) and signal ray at 805 nm (dashed green curve). (a) 1 mm long BBO crystal. (b) 3 mm long BBO crystal. (c) 6 mm long BBO crystal. (color online).

For the 3 mm crystal (Fig. 15b), the phase matching peaks for a signal at 805, 810 and 815 nm are shown to be of comparable width and separated enough that the overall angular width of the full spectrum is greater than the angular width of any one frequency. As the crystal length increases to 6 mm (Fig. 15c), the angular width of any single frequency component narrows, while the overall width of the 10 nm band is unchanged, as this width is set by the dispersion constants of the BBO, not the crystal length. Conversely, when the crystal length is reduced to 1 mm (Fig. 15a) the width of a single frequency component increases, improving the spatial resolution available to a single frequency component.

We make note of two aspects of the point spread functions illustrated in Figs 14b and c for the single frequency and slow detector cases – the square shape and the offset to one side of the center of the idler slit. Both of these arise from the linear dependence of the phase mismatch on the transverse k vector, κ, in Eq. 9, which is a characteristic of Type II phase matching. By integrating over $x_{os}$ the single frequency degenerate PSF of Eq. 11 can be rewritten as

$$CCR(0, z_{1i}, x_{2s}, d_2) = \left| \int d\kappa_i \ sinc\left(\frac{\Delta kL}{2}\right) e^{\frac{i\Delta kL}{2}} e^{-i\kappa_i x_{os}} e^{i\frac{\lambda_i d_2}{4\pi}\kappa_i^2} \exp\left[ -\frac{\left(\frac{\pi x_{2s}}{\lambda_s d_2}+\frac{\kappa_i}{2}\right)^2 \left(\left(\frac{\lambda_s d_2}{\pi a_p}\right)^2 + \frac{i\lambda_s d_2}{\pi}\right)}{\left(\left(\frac{\lambda_s d_2}{\pi a_p^2}\right)^2+1\right)} \right] \right|^2 \quad (17)$$

For the conditions of our experiments

$$\frac{\lambda_s d_2}{\pi a_p} \ll 1 \quad (18)$$

With this approximation, the important terms in Eq. 17 can be regrouped to give

$$CCR(0, z_{1i}, x_{2s}, d_2) = \left| e^{-\frac{x_{2s}^2}{a_p^2}} \int d\kappa_i \, sinc\left(\frac{\Delta k L}{2}\right) e^{\frac{i\Delta k L}{2}} e^{-\frac{a_p^2 \kappa_i^2}{4}\left(\frac{\lambda_s d_2}{\pi a_p^2}\right)^2} e^{-i\kappa_i x_{2s}} \right|^2 \quad (19)$$

The PSF can now be seen to be a Fourier transform of the product of a phase matching term $sinc\left(\Delta k L/2\right) e^{i\Delta k L/2}$ and a term that describes the influence of the numerical aperture imposed by the pump beam $exp\left[-\frac{a_p^2 \kappa_i^2}{4}\left(\frac{\lambda_s d_2}{\pi a_p^2}\right)^2\right]$. When the phase matching acceptance angle is so narrow that it dominates the range of spatial frequencies available for imaging, the PSF is effectively the Fourier transform of a sinc function. This gives the square profile in Figs. 14b and c for the longer crystals. The offset from the center of the slit arises from the phase term $e^{\frac{i\Delta k L}{2}}$. Physically this is related to birefringent walk off between the signal and idler photons. Signal and idler pairs that are generated near the exit of the crystal will remain relatively close to each other, while those generated near the entrance of the crystal will separate with the extraordinarily polarized photon propagating to one side of center. Using the expression in Eq. 10, the center of the PSF is seen to be

$$x_{2s,c} = -\frac{1}{n_s^e(\Delta\omega,\theta)} \frac{\partial n_s^e(\Delta\omega,\theta)}{\partial \theta} \frac{L}{2}. \quad (20)$$

This is exactly the offset expected from standard birefringent walkoff originating from the center of a crystal of length L.

Each of these effects arises from the linear dependence of $\Delta k$ on $\kappa$ in the argument of the sinc function and the exponential phase and is not present in type I phase matching. When the spread of spatial frequencies is determined by the numerical aperture of the pump beam as in the 1 mm crystal in Fig 13a, the PSF becomes the Fourier transform of a Gaussian function and is more bell-shaped. However, the offset from center is still apparent.

## 4. DISCUSSION.

The theoretical and experimental results shown here indicate the spatial and temporal resolution achievable in entangled ghost imaging depend on a combination of experimental parameters, phase matching properties of the SPDC crystals and the relative value of the time response of the single photon detectors and the coherence time of the biphoton. We have shown that the resolution obtained for type II down conversion is, in general, highly astigmatic with much smaller (worse) resolution in the e-direction than the o-direction. The SPDC interaction forms a frequency selective spatial filter at the SPDC crystal in addition to an aperture stop caused by the diameter of the pump beam, limiting the spatial resolution for slow detectors. It is this effect that resulted in the discrepancy of our experimental measurements and the back-propagation model of ref. 5. In that model, all frequency components acquire the necessary spread of spatial frequencies independently and therefore can form a high-resolution image even when combined incoherently. In the experimental situation illustrated here, a high-resolution image is obtained only if the spatial frequencies of different components can be combined coherently, which requires a detector that is faster than the biphoton coherence time.

We also might consider the physics involved in the relation between the temporal coherence of detectors that have no wavelength information and the spatial resolution of the correlated image. When viewed within the photon picture, we can note that the signal and idler photons are temporally coincident only to within the biphoton coherence time. A slow detector will consider that all signal idler photon pairs that are detected within the biphoton coherence time are coincident, and all such pairs will contribute to the resulting correlated image. A fast detector however will consider that signal-idler pairs that are separated by more than the detector resolution time, but less than the biphoton coherence time, as non-coincident. Such pairs will not contribute to the image obtained with fast detectors. As a result, the fast and slow detectors will sample different photon sub-ensembles.

Such reasoning can motivate why the images obtained with fast and slow detectors might be different, but we need to go further to relate detector resolution time to image spatial resolution. To see this we consider the signal and idler from a wave amplitude picture. In the classical limit this can be considered as classical field amplitudes, while at the quantum level they would be considered as probability field amplitudes. The temporal amplitude distribution of a polychromatic signal can be represented at the sum over temporal modes each with different random phases:

$$E_{i,s}(t) = \sum f_{ni,s} e^{-in\delta\omega_{i,s}t} e^{i\varphi_{i,s}(n\delta\omega_{i,s})} \tag{21}$$

where $f_{ni,s}$ is a mode amplitude and $\varphi_{i,s}(n\delta\omega_{i,s})$ is the phase of the individual idler or signal components that is randomly distributed over $2\pi$. The temporal envelope $E_{i,s}(t)$ typically consists of a low amplitude, randomly varying signal where the various wave components are not in phase with each other, punctuated by a few narrow, higher peaks where more of the components are in phase. We might then be encouraged to consider that the probability of detection of a signal idler pair is higher at the peaks where the components are in phase than at the lower amplitudes when the mode components are dephased. As a result, it might be expected that the single photon detection is biased toward detection when the signal and idler modes are in phase. Use of a fast detector would then ensure that the detection is over before the modes have a chance to dephase.

Further consideration reveals, however, that because the tall peaks when the modes are phased together are relatively sparse in a broad band signal, the total probability for detection of a single event with dephased modes is as great or greater than detection of an event with phased modes. In such a situation, even though the phases of the different modes are frozen in time during the detection by the fast detector, the image they form will be of low resolution when the modes are dephased and the individual spectral frequency components can be associated with different bands of spatial frequency as in Figs. 15b and c.

With regard to combining the modes at different angles and different wavelengths coherently in the image, we can take note that detectors that are faster than the coherence time of the biphoton would not be capable of distinguishing the difference in frequency among the different signal modes at differing wavelengths. However, the individual modes are born with phases that are randomly distributed relative to each other, and the fast detector would freeze the random phase distribution in time during the measurement. Thus, even though the difference in frequency might be made inconsequential by the fast detector, the modes with different wavelengths and therefore different spatial frequency bands would still carry randomly distributed phases and would not, by themselves, combine to form a single narrow PSF.

However, we then take note that the image formed in the ghost imaging interaction is a correlated image involving the product of the signal and idler creation operators in the form $a_s^\dagger(\kappa_s, \lambda_s) a_i^\dagger(\kappa_i, \lambda_i) e^{i(\varphi_s(\lambda_s) + \varphi_i(\lambda_i))}$. Even though the signal and idler phases of different spectral modes are randomly phased with respect to each other, the phases in the exponent of the product add to the pump phase, which is taken to be constant here, for each individual signal-idler wavelength pair:

$$\varphi_s(\lambda_s) + \varphi_i(\lambda_i) = \varphi_p \tag{22}$$

As a result, the random signal and idler phase terms can be removed from the frequency integrals of Eq. 16. Thus, when the fast detector is used, the different frequency amplitudes combine coherently in phase as well as frequency and, when the different frequency components are associated with different spatial frequency bands, as in Figs 11b and c, the different spatial frequency bands combine coherently, forming an image with higher spatial resolution.

ACKNOWLEDGEMENTS. This work was supported by the Office of Naval Research.